\documentclass[final]{jfm}

\usepackage{graphicx}
\usepackage{newtxtext}
\usepackage{newtxmath}
\usepackage{natbib}
\usepackage{hyperref}
\hypersetup{
    colorlinks = true,
    urlcolor   = blue,
    citecolor  = black,
}
\newtheorem{lemma}{Lemma}
\newtheorem{corollary}{Corollary}
\newcommand{\RomanNumeralCaps}[1]
\linenumbers

\renewcommand{\vec}[1]{\boldsymbol{#1}}

\usepackage[dvipsnames]{xcolor}
\newcommand{\AW}[1]{#1} %{\textcolor{NavyBlue}{#1}}
\newcommand{\MB}[1]{#1} %{\textcolor{NavyBlue}{#1}}

\title{Longitudinal vortices in unsteady Taylor--Couette flow: solution to a 60-year-old mystery}

\author{Ashley P.\ Willis\aff{1}
  \corresp{\email{a.p.willis@sheffield.ac.uk}}
  \and Michael J. \ Burin\aff{2}
}

\affiliation{
\aff{1}Applied Mathematics, School of Mathematical and Physical Sciences, Hicks Building, University of Sheffield,
Sheffield S3\,7RH, UK
\aff{2}Department of Physics, California State University, San Marcos, California 92096 USA}

\begin{document}
\maketitle

\begin{abstract}
%
%\AW{[TEST AW]}  
%\MB{[TEST MB]}
%
%  \AW{[Change $r$ to $r-R_i$ in all plots?]}
%
Applying a sufficiently rapid start-stop to the outer cylinder of the Couette--Taylor
system, structures approximately aligned with the axis were recorded in the classic work of 
\cite{coles1965transition}.
These short-lived rolls are oriented perpendicular to the classic Taylor-vortex rolls.
In this work we report numerical observation of this instability, guided by a more recent experimental observation.
The instability is shown to be related to an inflection in the azimuthal velocity profile, a finding consistent with the experimental observations of its emergence during the deceleration phase. 
Despite the transient nature of start-stop experiments, we show that the instability can be linked to that of the oscillating boundary layer problem of Stokes.  
There are several reasons why the instability may have remained elusive, both for experimental observation and \MB{for} the idealized system.  We look in more detail at dependence on the radius ratio for the Taylor--Couette system and find that,
in the case where the size of the rolls scales with the gap width,
for radius ratios any lower than that used by Coles, $R_i/R_o=0.874$, the instability is quickly overrun by axisymmetric rolls of G\"ortler type.
\end{abstract}

\begin{keywords}
%Authors should not enter keywords on the manuscript, as these must be chosen by the author during the online submission process and will then be added during the typesetting process (see \href{https://www.cambridge.org/core/journals/journal-of-fluid-mechanics/information/list-of-keywords}{Keyword PDF} for the full list).  Other classifications will be added at the same time.
\end{keywords}

%{\bf MSC Codes }  {\it(Optional)} Please enter your MSC Codes here

\section{Introduction}
\label{sec:intro}

The nature of the flow of fluid between two rotating cylinders has long played a special role in the study of fluid dynamics and dynamical systems.  \cite{taylor1923viii} demonstrated that linear stability analysis, using the no-slip condition, could accurately predict the onset of rolls.  A large number of distinct flow regimes were soon identified, and \cite{coles1965transition}(figure 2 therein) created a summary graphic in the plane defined by the rotation rates of the inner and outer cylinders. This plot included Taylor's original stability line, above which greater rotation of the inner cylinder rendered the flow unstable to `singly periodic flow' (Taylor-vortices), and catalogued regions of `doubly periodic flow' (wavy Taylor-vortices), \MB{in addition to }spiral and turbulent flows.  Coles was particularly interested in the multiplicity of wavy Taylor vortex states, of various azimuthal wavenumber, and transitions between them.  A celebrated version of this diagram, including a greater range of flows, was produced by \cite{andereck1986flow}.
It is now a little over 100 years since Taylor's landmark paper, 60 years since the classic work of Coles, and the rich dynamical landscape of the Taylor--Couette flow, for the steady rotating case alone, is well recognised.

% \subsection{Unsteady Taylor--Couette flow}

Multiple flow states for the same rotation rates are not easily represented on a summary graphic, and figure 8 of \cite{coles1965transition} is a \MB{prime} example of this complexity. 
The addition of time-dependence likewise introduces an extra dimension of complexity, and expeditions in these directions are fewer but significant.  
G\"ortler rolls appear when flow encounters a curved surface, and like Taylor-vortex rolls, they are aligned with the direction of the flow and caused by a centrifugal instability.
G\"ortler rolls have been observed to form and eventually evolve into Taylor-vortices for the case of an impulsive start of the inner cylinder, see e.g. \cite{neitzel1995transient}. 
Studies of an impulsive stop for the outer cylinder include   
\cite{kohuth1988experiments,verschoof2016self,ostilla2017life,singh2021turbulence}, where rolls of many sizes appear and decay, leaving behind the largest \MB{scale}.

% \subsection{modulations}

Periodic modulations of the rotation rate of either cylinder have been studied. For example, \cite{ern1999flow} found regimes of both destabilisation and stabilisation, while \cite{youd2003reversing} observed that Taylor-vortex rolls may or may-not reverse the direction of the rolls. Axial oscillations of the inner cylinder have been shown to suppress Taylor instability by \cite{marques1997taylor}. In the context of planetary libration, an interesting set of experiments and calculations with modulated rotation rates have been performed, e.g.\ \cite{noir2010experimental,lopez2011instabilities}. In all cases discussed above, the same instabilities are manifest, i.e.\ the appearance of modified Taylor-vortices or G\"ortler rolls.
%
% \subsection{modturb}
\MB{
Other related studies have focused on the effect of surface modulation on a turbulent background state, e.g.\ \cite{verschoof2018periodically} applied oscillations of the inner cylinder in high Reynolds number experiments and observed how the flow responded to different oscillation rates.}
%$\$
%Rei = 100,000, 
%observed that
%for shorter modulation periods,
%the flow responds with a phase delay.
%effective turbulent mixing  prevented a radially-dependent response.

%Numerical work on planar Couette flow at Re = 30,000 was performed by \cite{akhtar2022effect} under rotation and with a 10\% modulating amplitude, revealing a significant difference of system response between cyclonic and anticyclonic states. Further work by \cite{akhtar2024super} at the same Re, but without rotation, noted that the response within the turbulent flow can be well approximated by a laminar model as long as the perturbation amplitude does not exceed the base flow.}   

%\AW{[Add citations suggested following ICTW Barcelona meeting:\\
%Reference from Alfredo - \\
%\cite{ern1999flow} \\
%References from Rudolfo - \\
%Periodic modulation experiment:
%\cite{verschoof2018periodically}\\
%Periodic modulation DNS on PCF:
%\cite{akhtar2022effect}\\
%Sudden stop experiment at twente:
%\cite{verschoof2016self} \\
%Sudden stop simulations:
%\cite{ostilla2017life} \\
%]}

\subsection {Experimental observations of longitudinal vortices}

\begin{figure}
  \centerline{
    \begin{picture}(0,0)
  \put(-10,220){({\it a})}
  \put(175,220){({\it b})}
  \end{picture} 
  \includegraphics[height=75mm]{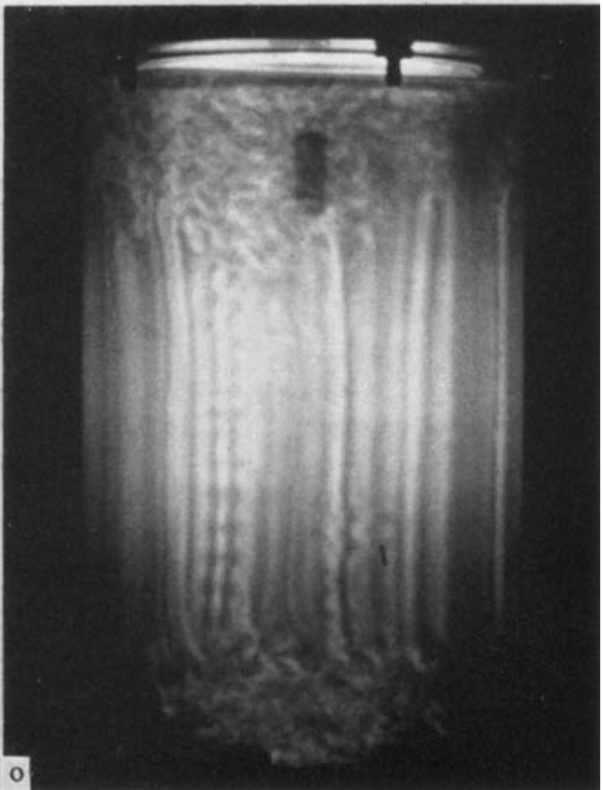}
  ~~~~~~~
  \includegraphics[height=75mm]{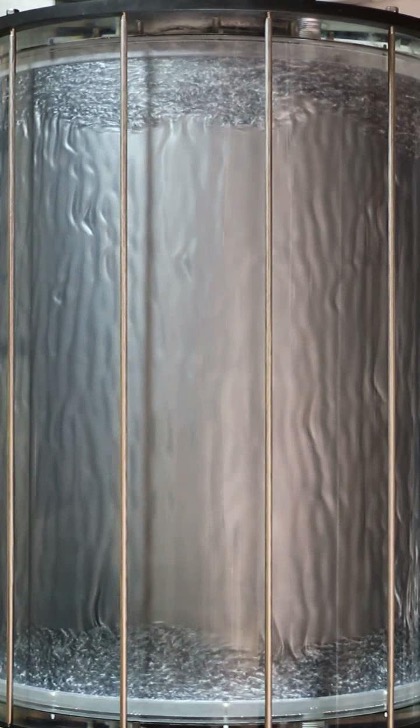}
  }
  \caption{
  \label{fig:expts}
     ({\em a}) Figure 22(o) of
     \cite{coles1965transition}, captioned `Instability following start-stop motion of outer cylinder.'
     ({\em b}) Similar observation from the experiments of \cite{Burin12}
  }
\end{figure}

\cite{coles1965transition} includes an image of a flow, figure 22(o) therein and 
reproduced in figure \ref{fig:expts}({\em a}) here, 
that appears unlike any of those mentioned above.  While those flows are dominated by azimuthally aligned rolls that wrap around the cylinder, here the waves are aligned with the axis.
In a footnote (p410), Coles describes 
his method for establishing wavy Taylor vortex flow with a high `tangential' wavenumber, now more
commonly referred to as the azimuthal wavenumber $m$: 
{\it ``The usual experimental method for establishing a flow with a high tangential wavenumber was to approach the final operating condition from the singly-periodic side} [from axisymmetric Taylor rolls]
{\it with two cylinders initially rotating in the same direction. A sufficiently fast stop for the
outer cylinder then usually led to a flow with a large number of tangential waves, probably
as a result of Tollmien instability in the unsteady viscous layer on the wall of the outer
cylinder (cf.\ figure 22(o), plate 12).''}
The Tollmien--Schlichting instability is a viscous instability that can occur in a boundary layer flow, such as when a unidirectional flow encounters a flat plate.  The rolls that arise within the boundary layer are aligned spanwise to the direction of the flow.
In principle, it is possible the sudden-stop could indeed set up a boundary-layer type flow and hence Tollmien instability.  However, Coles' caption to figure 22(o) reads only 
{\it ``(o) Instability following start-stop motion of outer cylinder.''}
The sudden-stop and start-stop experiments are potentially quite different!  Unfortunately, no parameters for the start-stop experiment were provided.

More recently, during an experimental campaign to characterize turbulence driven by outer cylinder motion \citep{Burin12}, a brief set of start-stop observations were made, revealing what appears to be the same azimuthal instability as described above. See figure \ref{fig:expts}(b). The instability in this \MB{case} was observed to arise during the decelerating phase, and for only a narrow gap width ($d=6.3$ mm and radius ratio $R_i/R_o=0.97$), although its apparent absence for wider gap widths could be due to visualization shortcomings.
A qualitative dependence of the instability wavelength with respect to forcing was observed, with initial acceleration provided by falling weights coupled to the drive system, while mechanical friction provided deceleration. As the acceleration increased, the wavelength increased by a factor of two: from about 1.3 to 2.5 cm, or (approximately) 2 to 4 gap widths. For smaller accelerations the instability was not seen, and for higher ones the turbulence emanating from the end-caps engulfed the central region before the instability could presumably appear.   
\MB{The maximum Reynolds number of the outer cylinder varied in these initial exploratory experiments, although the period and rate of both the acceleration and deceleration phases contain significant uncertainty due to the simple driving mechanism. Nonetheless, this second appearance of the instability in question did reveal that it occurs during the deceleration phase.}

To our knowledge, there has been no systematic study of this transitory TC flow state. 
In this paper, using new simulations alongside these 
scarce %\MB{occasional} 
laboratory observations, we aim to clarify when the longitudinal vortices should appear, with respect to the type of motion of the outer cylinder, and to identify the instability mechanism.  We show how this particular unsteady flow may be understood within the wider context of unsteady boundary flows, along the way linking Taylor--Coutte flow to a branch of fluid mechanics reaching back to Stokes’ oscillating boundary layer problem.

\section{Methods}

We consider the flow between concentric cylinders of radius $R_i$ and $R_o$,
where the subscripts indicate the inner and outer cylinders respectively, 
with rotation rates $\Omega_i$ and $\Omega_o$.  The radius ratio is denoted
by $\eta=R_i/R_o$ and gap width $d=R_o-R_i$.
Scales used in the non-dimensionalisation of the governing equations are
the gap width $d$ and the viscous diffusion timescale $d^2/\nu$, where
$\nu$ is the kinematic viscosity, which combine to give a velocity scale $\nu/d$.
Quantities in all figures are non-dimensionalised with these scales.

The dimensionless Navier--Stokes equations are
\begin{equation}
    \label{eq:NS}
    \partial_t\vec{u}+\vec{u}\cdot\bnabla\vec{u} = -\bnabla p+\nabla^2\vec{u}\, ,
    \quad \bnabla \cdot \vec{u} = 0 \, .
\end{equation}
Note that the Reynolds numbers, $\Rey_{i,o}=R_{i,o}\Omega_{i,o}d/\nu$,
do not appear in the governing equation but determine the speed of the boundaries:
$u_\theta=\Rey_i$ 
and $u_\theta=\Rey_o$
at dimensionless radii 
$r_i=R_i/d=\eta/(1-\eta)$ and $r_o=R_o/d=r_i+1$
respectively.

When $\Rey_{i,o}$ are constant, steady circular-Couette flow
is given by
\begin{equation}
\label{eq:CCF}
   \vec{u} = V_0(r)\hat{\vec{\theta}},
   \qquad
   V_0(r) = \frac1{1+\eta}
   \left[
      ( \Rey_o - \eta \, \Rey_i) r + \frac{\eta}{(1-\eta)^2}
      ( \Rey_i - \eta \, \Rey_o) \frac1{r}
   \right] .
\end{equation}
To implement the case when time-dependent $\Omega_{i,o}(t)$ imply time-dependent 
$\Rey_{i,o}(t)$,
we consider the deviation $\vec{u}'$ of the velocity field from (\ref{eq:CCF}) evaluated 
at each instant,  
$V_0(r,t)=V_0(r;\Rey_i(t),\Rey_o(t))$, i.e.
\begin{equation}
\label{eq:decomp}
    \vec{u}(\vec{x},t) = V_0(r,t)\hat{\vec{\theta}} + \vec{u}'(\vec{x},t) \, .
\end{equation}
The field $V_0(r,t)$ is used only in the computations and should be distinguished from the `mean profile' $V(r,t)$, which denotes the
azimuthal component of the 
$\theta$- and $z$-average of the total flow field $\vec{u}$.
\AW{
The base component of the flow
$\vec{u} = V_0\,\hat{\vec{\theta}}$
satisfies 
%$\vec{u}\cdot\bnabla\vec{u}=\vec{0}$ and 
$\nabla^2\vec{u}=\vec{0}$.
Substituting (\ref{eq:decomp}) into (\ref{eq:NS}) and using 
this property,
%these properties, 
the} 
governing equation for the deviation field $\vec{u}'$ can be written
\begin{equation}
    \label{eq:NSint}
    (\partial_t-\nabla^2)\vec{u}'
    = \vec{u}\times(\bnabla\times\vec{u}) 
      - \bnabla p
      - (\partial_t V_0) \,\hat{\vec{\theta}} 
      \, ,
\end{equation}
subject to $\bnabla\cdot\vec{u}'=0$
with boundary conditions $\vec{u}'=\vec{0}$ at $r=r_{i,o}$.
\AW{A vector calculus identity has been applied to the advection term to give
the rotational form of the governing equation, with the double-cross term on the right.  This is slightly easier and less expensive to evaluate numerically.}

The code used for time integration of (\ref{eq:NSint})
is a long-standing variant of the {\em Openpipeflow}  solver
\citep{willis2017openpipeflow}, and modification for this case required only the addition of the forcing term, i.e.\ the known last term on the right-hand side
of (\ref{eq:NSint}).
Periodicity is assumed 
in the axial dimension over a length $2\pi/\alpha$,
and $m$-fold periodicity can be imposed in the azimuthal dimension.
Variables are discretised 
\AW{via a double-Fourier decomposition, and finite differences are used in the radial dimension:}
%in the form 
\begin{equation}
   \label{eq:Fourier}
   a(r_n,\theta,z) =
      \sum_{|k|<K} \sum_{|m'|<M} 
      A_{km}(r_n) \, \mathrm{e}^{\mathrm{i} (\alpha kz + m\,m'\theta)},
      \quad
      n = 1,\dots N \,.
\end{equation}
\AW{Each Fourier mode $A_{km}$ is evaluated at the radial points
$r_n$, which are located at the $N$ extrema the Tchebyshev polynomial $T^{N-1}(x)$ defined on $x=r-r_i \in [0,1]$.  These points are bunched towards the walls within the boundary layers.}
$N=100$ for most calculations, while, for linear stability calculations, 
$K$ and $M$ are only as large as required to capture the first few modes.
The Crank--Nicolson method is applied to the diffusion term, while 
an Euler-predictor and Crank--Nicolson corrector are applied 
the first and last terms on the right-hand side of (\ref{eq:NSint}).   Via the pressure-Poisson equation, $p$ is used to project the 
velocity onto the space of divergence-free functions, using an influence-matrix
technique to implement the no-slip and divergence-free condition at the wall
simultaneously.
The difference between the predictor and corrector may be used to control the timestep,  but for the calculations here, the timestep is fixed at $\Delta t=10^{-6}$.  
(Note that the dimensionless advective timescale $1/|\vec{u}|\sim 1/\Rey_o$.)

The mean profile ($\theta,z$-average) 
is the component
%\begin{equation}
  $  V(r,t) = u'_{\theta,00} + V_0(r,t)  $,
%\end{equation}
where the subscript 00 indicates the $k=m'=0$ mode in (\ref{eq:Fourier}).
In the absence of perturbations, the mean profile is quickly computed 
by simulating with $K=M=1$.
Eigenvalues, which give the growth rates of linear perturbations to the base flow,
are calculated by combining time-integration with 
Arnoldi iteration.
\AW{
This method applies repeated re-orthogonalisation to the perturbation
via the Gram--Schmidt process,
which accelerates convergence 
to the leading mode, and enables
capture of further eigenmodes and eigenvalues.  For details, see e.g.\ \cite{chu2024minimal} \S 2.2\,.
}

\section{Results}

\subsection{Approximation of experiment}
\label{sect:ApproxOfExpt}

%In this non-dimensionalisation, $\Rey_1$ and $\Rey_2$ correspond to the 
%speed of the inner and outer cylinders at $R_1$ and $R_2$.  

\AW{
The recent experimental observations of longitudinal vortices for a start-stop of the outer cylinder provided the following parameter configuration, which is later rationalised in \S\ref{sect:stokesBL}.
An example video may be found in the supplementary material; 1 second in the video corresponds to approximately 0.024 viscous times $d^2/\nu$.}

\AW{
Instability was observed for the narrow gap, $\eta=0.97$,
and systematic experiments used a weight-and-pully system to start the motion, implying approximately constant acceleration.  The outer cylinder was also observed to come to an abrupt halt, suggesting use of a constant decelleration until no motion is reached, rather than say an exponential slow down.
Therefore,}
%The inner cylinder is fixed, $\Rey_i=0$,
%and supposing that 
%instability in the preliminary experiment was excited by an approximately constant acceleration and deceleration, 
we begin with a linear ramp up and down 
in the Reynolds number for the outer cylinder
\begin{equation}
  \Rey_o(t) = 
  \left\{ \begin{array}{ll}
     U_1\,{t}/{T_1},  & 0 \le t < T_1, \\
     U_1\,{(T_2-t)}/{(T_2-T_1)}, & T_1 \le t < T_2, \\
     0, & T_2 \le t \, ,
  \end{array}\right.
\end{equation}
with experimentally-estimated dimensionless parameters 
$U_1=8000$, $T_1=0.01$, $T_2=0.05$.
The inner cylinder is fixed, i.e.\ $\Rey_i=0$.
%Inspection of videos of the experiments 
%\AW{[Include in Supplementary Material?]} 
%for the narrow gap case, $\eta=0.97$,
%suggested that instability 
Instability was most frequently observed 
with an azimuthal wavelength $\lambda\approx 3.5$
(gap widths).  
  Using
$\lambda\approx 2\pi r_o/m$, where $r_o=1/(1-\eta)$
and $m$ is the rotational symmetry (see (\ref{eq:Fourier})), implied setting $m=60$.

\begin{figure}
  \centerline{
  \begin{picture}(0,0)
  \put(10,140){({\it a})}
  \put(220,140){({\it b})}
  \end{picture}
  \includegraphics[width=0.55 \textwidth]{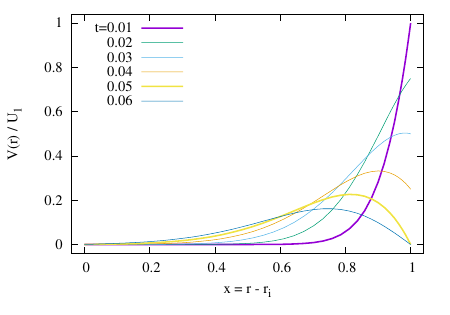}  \includegraphics[width=0.52\textwidth,trim=0 {-3mm} 0 0]{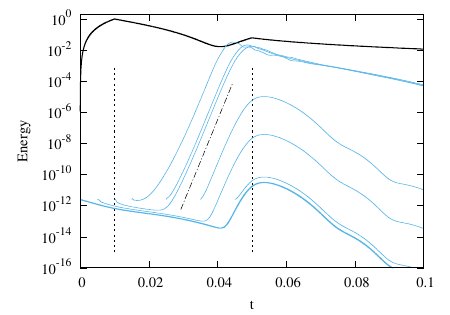}}
  \caption{
  \label{fig:prof}
  (\textit{a}) 
  Mean profile during the ramp up of the outer cylinder (from $t=0$ to $T_1=0.01$)
  and ramp down of the outer cylinder 
  (from $t=T_1$ to $T_2=0.05$).
  \\   (\textit{b}) \AW{
  Normalised energy of the mean flow,
  $E_{m=0}(t)/E_{m=0}(T_1)$ (black) and 
  energy of non-axisymmetric ($m\ne 0$) perturbations, 
  $E_{m\ne 0}(t)/E_{m=0}(T_1)$ (blue).
  Each blue line shows the development of a perturbation introduced at a different time, where the line starts. 
  Vertical lines indicate the times $T_1$ and $T_2$ and the slope indicates the 
  growth rate for the wavelength $\lambda\approx 3.5$ at $t=0.04$ of figure \ref{fig:Eigvals}.  }
  }
\end{figure}

Figure \ref{fig:prof}({\it a}) shows calculations of the azimuthal mean flow profile
 $V(r,t)$ during the ramp up and ramp
down of the outer cylinder.  
\AW{
The viscous time-scale $d^2/\nu$ was used for the non-dimensionalisation. 
Similarly, a time scale for the ramp-up stage may be written $\delta^2/\nu$, where $\delta$ measures how far the change in velocity of the outer boundary is expected to penetrate into the flow.
Expressing this in the timescale used for the non-dimensionalisation, we have 
%A time-dependent viscous length scale $\delta$ for the ramp-up stage, measuring the distance change in velocity of the outer boundary is expected to penetrate into the flow, 
%may be defined similarly via
%$t^* = \delta^2/\nu$, i.e.\ $\delta=\sqrt{\nu t^*}$.  Non-dimensionalising $t^*$ gives $t^*=
$\delta^2/\nu = t\,d^2/\nu$, which rearranges to $\delta/d=\sqrt{t}$.  
This implies that by the end of the ramp-up stage, at time $t=T_1=0.01$, non-zero velocity is expected to penetrate approximately one tenth of the way across the gap, in good agreement with the 
curve for $t=0.01$ in figure \ref{fig:prof}({\it a}).  From this time on, the ramp-down is seen in the velocity at the outer boundary, leaving behind a non-zero velocity profile in the interior.}

Velocity perturbations are added to $V(r,t)$ that are very small, so that the Fourier modes of the perturbation may be considered to be linearly independent.  
%As the rolls to be identified are axially aligned, 
\AW{For the case of axially-aligned rolls,}
we consider only modes with $k=0$ in (\ref{eq:Fourier}) so that the calculations are two-dimensional (axially-independent).  Small perturbations are added for $m'$ up to $7$, but the $m'=1$ mode is always found to be dominant. 
Figure \ref{fig:prof}({\it b}) shows time series of the energy summed over all azimuthally-dependent ($m'\ne 0$) perturbations.  
It was found that perturbations added at $t=0$ would not grow substantially,
\AW{such as the bottom-most curve}.
Instead, perturbations added at later times during the development of the mean profile were necessary to trigger transient instability.  (In the physical experiment, it is perhaps 
reasonable that some noise is always present, from end conditions at least.)
\AW{Where a growing mode becomes established, the}
growth rates appear to be approximately constant 
for times 0.02-0.04, which is during
the decelerating phase, consistent with the appearance of the instability in the experiment.

\begin{figure}
  \centerline{
  \begin{picture}(0,0)
  \put(5,140){({\it a})}
  \put(215,140){({\it b})}
  \end{picture}
  \includegraphics{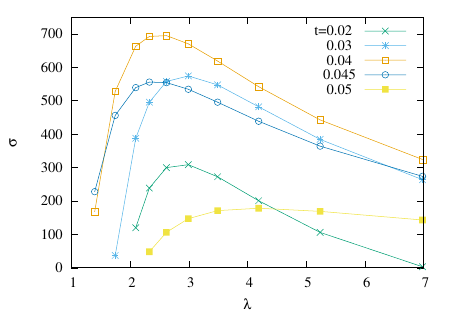}
  \includegraphics[trim={4mm} 0 0 0]{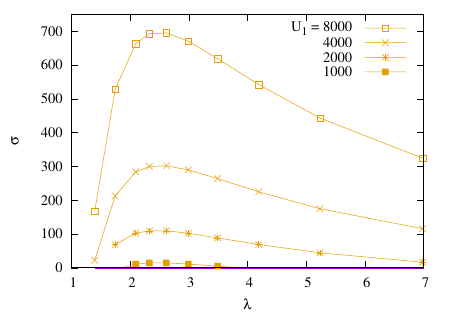}}
  \caption{
  \label{fig:Eigvals}
  (\textit{a}) 
  Growth rates $\sigma$ as a function of azimuthal wavelength $\lambda$, assuming frozen mean profiles
  at the given times.
  $U_1=8000,~T_1=0.01,~T_2=0.05$.
  (\textit{b}) 
  Growth rates at $t=0.04$ changing only $U_1$.
  }
\end{figure}

It is possible in our numerical calculations to 
freeze the mean profile $V(r,t)$ at a particular time
and to calculate the growth rates
of infinitesimal disturbances.  For this to 
produce physically reasonable growth rates, 
it assumes that there is time for the 
disturbances to grow substantially more rapidly
than changes in the base profile.  In figure 
\ref{fig:prof}({\em b}), it is seen that 
perturbations can grow many orders of magnitude
in a short time.
With this frozen profile assumption, 
figure \ref{fig:Eigvals}({\em a})
shows the growth rates of 
perturbations to the profile as a function of 
azimuthal wavelength $\lambda$ at several times.
Peak growth is observed at $t=0.04$ for a 
wavelength of $\lambda\approx 2.5$, but observe
that instability starts earlier at longer 
wavelengths, $\lambda\approx 3$ for $t=0.02$ (approximately when growth is first observed), which is a reasonable match to the wavelength observed in the experiment.

For the purely azimuthal base flow 
$\vec{u}=V(r,t)\hat{\vec{\theta}}$,
the nonlinear
terms in (\ref{eq:NS}) are zero, so that $\vec{u}$
can be multiplied through by an arbitrary constant and remains a solution.  
Therefore, if $U_1$ is changed without altering $T_1$ and $T_2$, the mean flow is multiplied by a factor
but does not change structure.  Looking at the time $t=0.04$ where growth was maximised, figure \ref{fig:Eigvals}({\em b}) shows that a change in amplitude of the start-stop changes the growth rate of perturbations, but has no noticeable effect on the preferred wavelength of instability.

\begin{figure}
  \centerline{
  \begin{picture}(0,0)
  \put(5,140){({\it a})}
  \put(185,140){({\it b})}
  \end{picture}
  \includegraphics[width=0.5\textwidth]{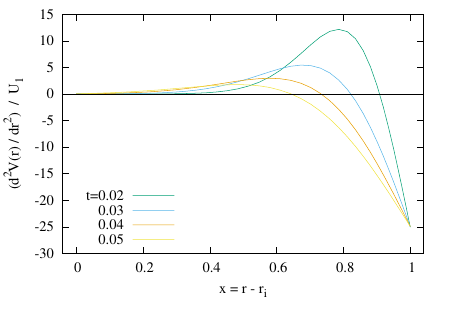}
  \vspace{4mm}
  \includegraphics[width=0.5\textwidth]{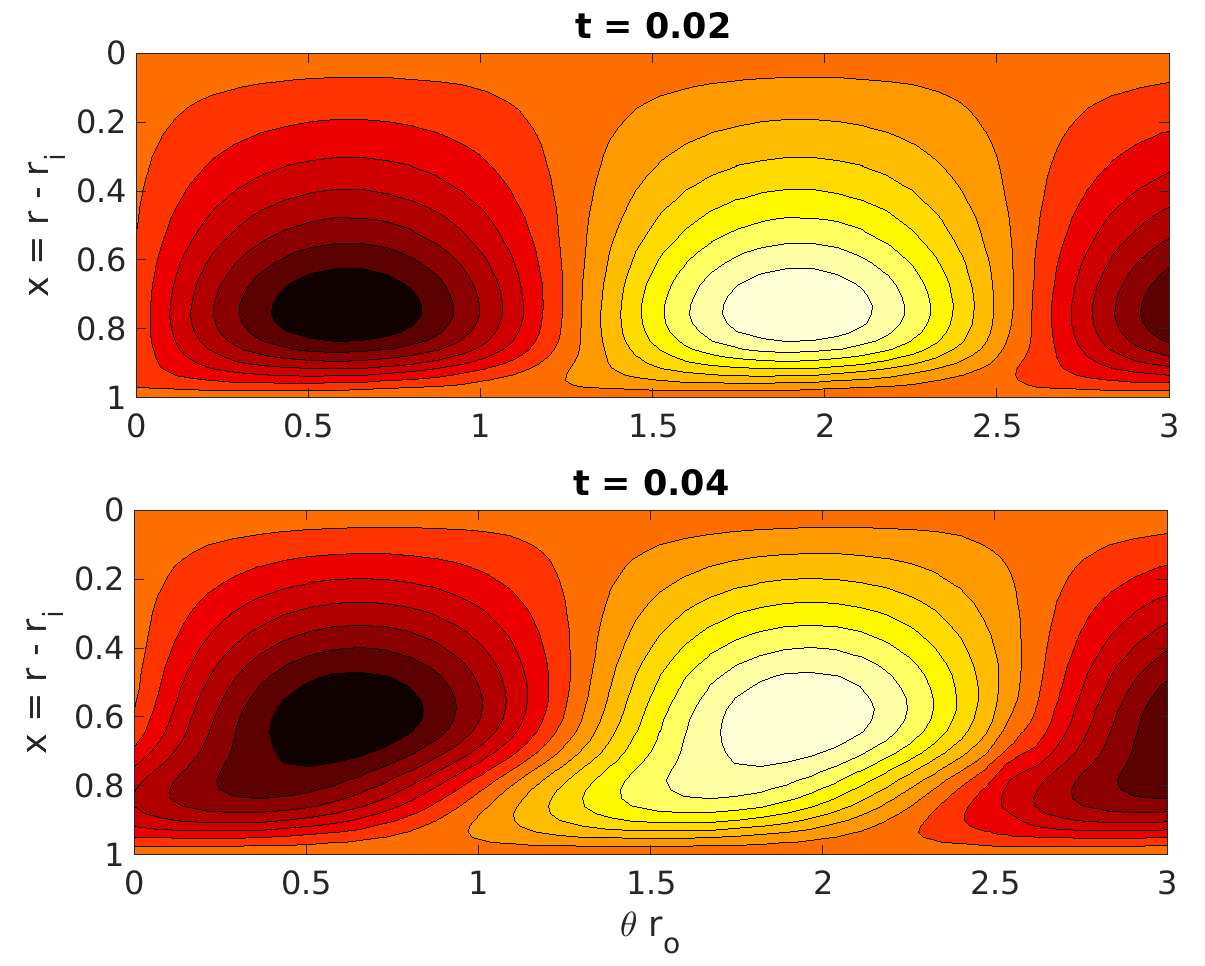}
  }
  \caption{
  \label{fig:prof2}
  (\textit{a}) 
  Second derivative of the mean profile \AW{for the parameters of figure \ref{fig:prof}} -- roots indicate the location of an inflection 
  point \AW{that appears strictly after $T_1=0.01$}. \\
  (\textit{b}) 
  Eigenfunctions of the leading unstable mode
  with
  $\lambda\approx 2.6$ at the given times; \AW{contours of streamfunction $\psi$ such that eigenfunction $\vec{u}'=\bnabla\times(\psi\hat{\vec{z}})$.}
  }
\end{figure}

\begin{figure}
  \centerline{
  \begin{picture}(0,0)
  \put(5,130){({\it a})}
  \put(205,130){({\it b})}
  \end{picture}
  \includegraphics[width=70mm]{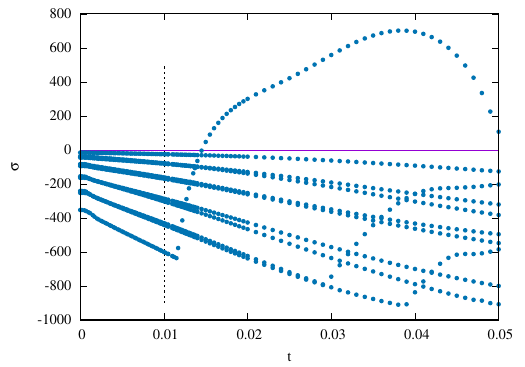}
  \vspace{4mm}
  \includegraphics[width=70mm]{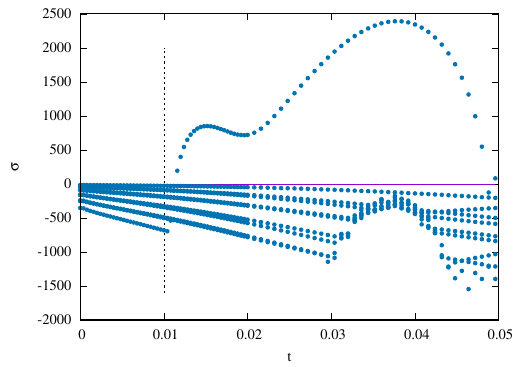}
  }
  \centerline{
  \begin{picture}(0,0)
  \put(5,130){({\it c})}
  \put(205,130){({\it d})}
  \end{picture}
  \includegraphics[width=70mm]{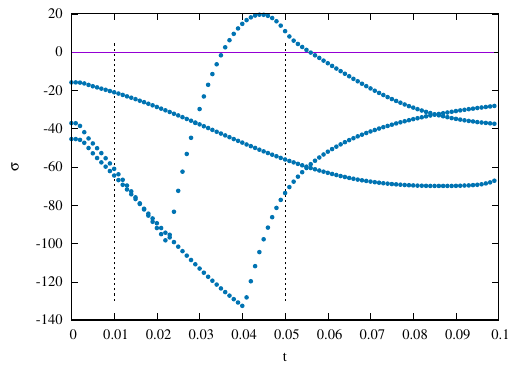}
  \vspace{4mm}
  \includegraphics[width=70mm]{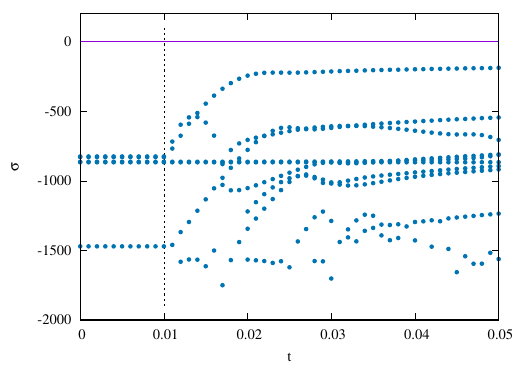}
  }
  \caption{
  \label{fig:Eigvals(t)}
  Leading growth rates $\sigma$ as a function of $t$ for $T_1=0.01$, $T_2=0.05$, $\lambda\approx 2.6$ $(m=80)$ with
  (\textit{a}) 
  $U_1=8000$, 
\AW{
  (\textit{b}) 
  $U_1=24\,000$,} and
  (\textit{c}) 
  for lower $U_1=1000$ shown over a longer period.
\AW{
 Vertical lines indicate the times $T_1$ and $T_2$.  Only one longitudinal mode goes unstable;
all other modes for $t\lesssim 0.03$ connect to the free-decay modes of $t=0$ when $\vec{u}=\vec{0}$.
  ~~(\textit{d}) 
  Growth rates for the ramp-down of a developed flow:
  $\Rey(t)=U_1=8000$ for 
  $-\infty<t<T_1$, then $\Rey(t)$ as before for $t>T_1$.
} }
\end{figure}

In the Introduction, it was noted that 
\cite{coles1965transition} had postulated 
that the rolls `probably' appeared as the result
of Tollmien instability, which appears in the 
boundary layer problem.  Acceleration of the boundary
sets up a velocity profile of the boundary layer type, but 
the decelerating phase introduces an inflection point, and it is during this phase that the instability is observed.
Figure \ref{fig:prof2}({\em a}) shows the second
derivative of the mean profiles at several times, so that the zero of this function 
tracks the position of an inflection point.
Using a wavelength $\lambda\approx 2.6$ ($m=80)$, 
around the value of theoretical preferred wavelength from figure \ref{fig:Eigvals},
figure \ref{fig:prof2}({\em b}) shows eigenfunctions at times
0.02 and 0.04 (as viewed from above, looking down the axis of the cylinders). 
%for $\lambda\approx 2.6$ ($m=80$).
Although the rolls are not centred precisely on the radial location of the inflection point, they are close and 
move further from the boundary as $t$ increases.
Figure \ref{fig:Eigvals(t)}({\it a}) shows the leading growth rates 
(of modes with strictly $m$-periodic dependence)
%(for modes with $m'=1$-dependence within the $m$-periodic domain)
as a function of $t$.  At $t=0$, the mean flow is zero, so all growth rates are 
associated with free decay modes.  
Modes associated with Tollmien instability should
arise during the ramp up from $t=0$ to $t=T_1=0.01$,
\AW{when the mean profile is similar to that for the boundary layer flow.}
We only see slow adjustment of the growth rates of the free-decay
modes, however, duing this period.  
From $t=T_1$, deceleration occurs, the inflection is introduced into the mean profile, and  
the mode associated with the observed longitudinal vortices quickly appears.

\AW{Using the viscous length scale $\delta$ introduced in relation to figure \ref{fig:prof}({\it a}), the accelerating phase may be compared with Stokes' problem for a flat plate impulsively accelerated to velocity $U_0$.  
A critical value for $\Rey_I=U_0\delta/\nu$  close to $\Rey_I=1485$ 
has been calculated by \cite{luchini2001linear}
for the appearance of Tollmien--Schlichting waves.  Using the relationship $\delta/d=\sqrt{t}$ and equating $U_0$ with $R_o\Omega_o$ gives
$\Rey_o=\Rey_I/\sqrt{t}$.  
Given that our boundary velocity is only ramped up to $U_1$, rather than impulsively started, this provides only a lower bound, $\Rey_o(t)\ge 1485/\sqrt{t}$,
where $\max(\Rey_o(t))=U_1$.
Within the ramp-up phase, the bound is lowest at $t=T_1=0.01$, implying an $\Rey_o$ greater than $14\,850$ would be required for Tollmien instability.  Calculations with $U_1$ three times larger at $U_1=24\,000$ showed growth rates  similar in form (figure \ref{fig:Eigvals(t)}{\it b}), still with no sign of instability before $T_1$ and instability arising immediately after $T_1$.  
Although Tollmien instability could in principle appear later $t$, it is unlikely since the velocity profile decays rapidly.  Moreover,
the appearance of instability at the same time adds extra weight to distinguish it from Tollmien instability, as,
when different $\Rey_o(t)$ are achieved,
the bound implies that such instability is expected to arise at a different time $t$. 
}

\AW{As the instability arises during the decelerating phase, it is reasonable to ask whether the accelerating phase is necessary.  Figure \ref{fig:Eigvals(t)}({\it d}) shows the growth rate for the case where the mean flow is developed with $\Rey(t)=U_1=8000$ for all $t$ up until $T_1=0.01$, then is followed by the same ramp down to $\Rey(t)=0$ at $t=T_2=0.05$.  All eigenvalues are negative, indicating that both the ramp up and ramp down phases are required for instability.
}

\subsection{Comparison with Stokes' oscillating boundary problem}
\label{sect:stokesBL}

Stokes found an analytic solution for the flow induced in a semi-infinite 
body of fluid by a sinusoidal oscillation of an infinite plate.  For a plate moving at (dimensional) speed $U=U_0\cos(\omega t)$, the viscous length scale for the time-varying boundary velocity is often defined
$\delta=\sqrt{2\nu/\omega}$.
%, which measures the penetration depth of time-varying boundary velocity perturbations.
Scaling with $U_0$ and $\delta$ gives a
Reynolds number $R^\delta=U_0\delta/\nu$.
\cite{von1974linear} considered the linear stability of the base flow 
at each instant for a channel of width $d$, where the second wall
is stationary.  For most calculations, the ratio of scales was $\beta=d/\delta=8$, for which it was shown that the influence of channel width was negligible.
The most unstable %dangerous 
velocity profile was found to be at time $t=\pi/2$
and that rolls of wavenumber $\alpha=0.5$ first go unstable at $R^\delta=86$.
%The corresponding wavelength is $4\pi\,\delta=(\pi/2)\,d$.  
\cite{blondeaux2021revisiting} have more recently revisited this analysis
for $\beta\to\infty$, confirming the results of \cite{von1974linear}.  They also examined for which $R^\delta$ there is instability in the accelerating phase or over
the whole cycle.

To produce the counterpart instability in the Taylor--Couette apparatus,
we suppose that the gap for the channel, $d=\beta\delta$, coincides with the Taylor--Couette gap width $d$, and 
expect that correspondence will be best for the 
narrow gap,
where curvature terms are small relative to the viscous terms.
%. \MB{due to reduced curvature, or just a nearby second wall?}
%As the ramp up/down experiments are characterised by a period rather than a frequency, we write 
For the Taylor--Couette system we put
$\Rey_o=U_1\sin(2\pi t/T)$.  

\begin{figure}

  \centerline{
  \begin{picture}(0,0)
  \put(5,140){({\it a})}
  \put(200,140){({\it b})}
  \end{picture}
  \includegraphics[width=0.52\textwidth]{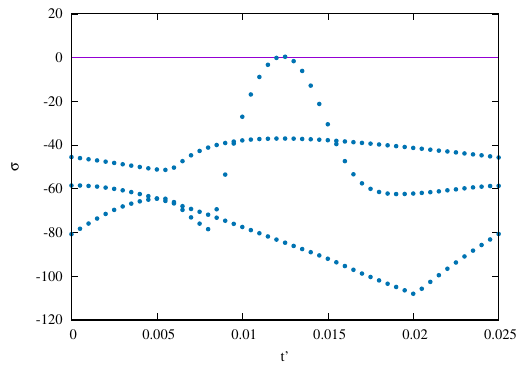}
  ~~~
\includegraphics[width=0.53\textwidth,trim=0 0 {-10mm} 0]{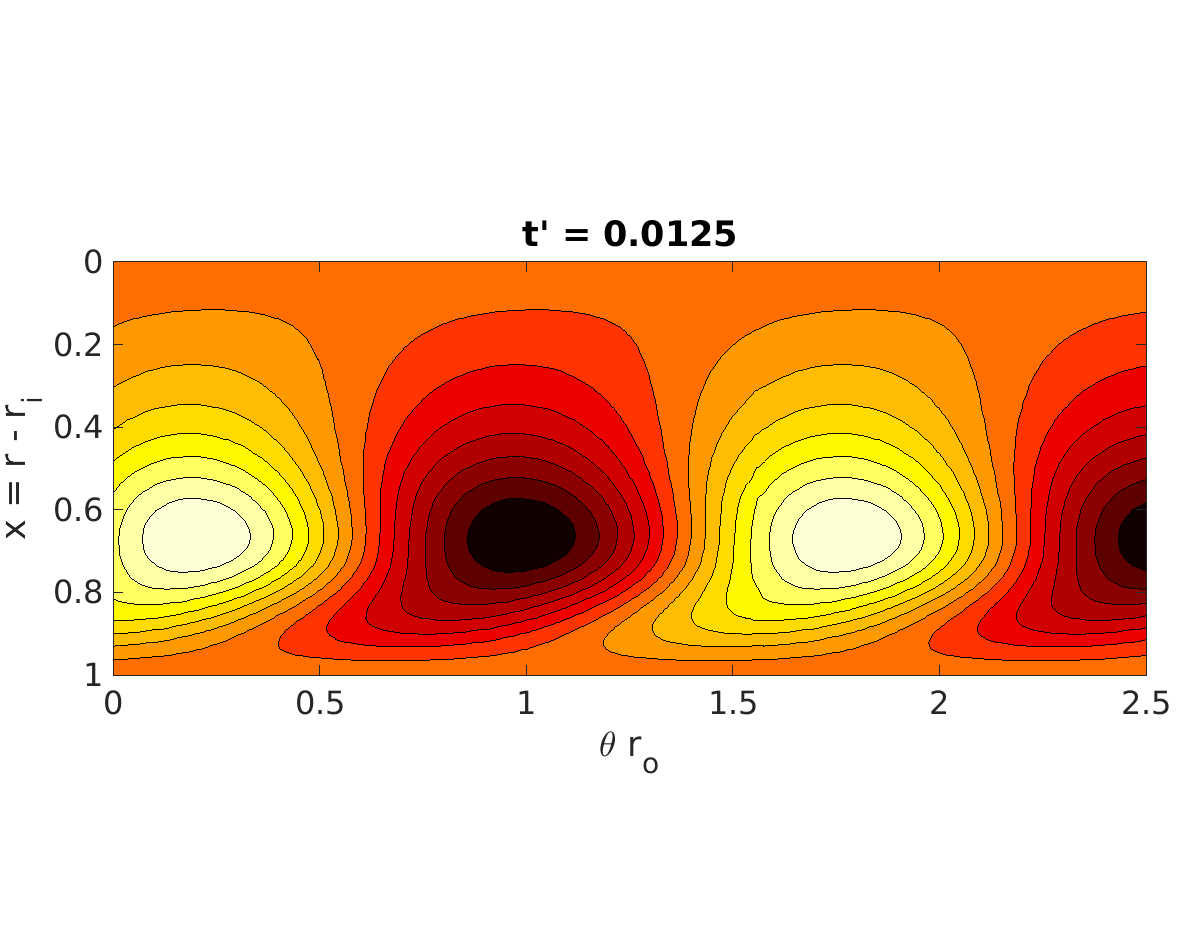}
  }
  \caption{
  \label{fig:KD}
  ({\it a})
  Three leading growth rates 
  for a sinusoidal oscillation of period $T$ as a function of $t'=t-9.25\,T$
  in the Taylor-Couette configuration.  Parameters for critical $U_1$ and wavenumber $m$ predicted from Stokes oscillation in a channel.  Calculations cover
  half of the period $T=0.05$, as the base flow reverses in the second half
  \AW{and eigenvalues repeat.  $Re_o(t)$ is maximum at $t'=0$ and is zero at $t'=T/4=0.0125$.
  ({\it b})
  Eigenfunction of the neutral mode.
  }
  }
\end{figure}

Converting  scales from the 
channel to Taylor--Couette system,
we have
$\delta=\sqrt{2\nu/\omega}=\sqrt{2\nu(T(d^2/\nu)/(2\pi))} = d\,\sqrt{T/\pi}$, then using
$\beta=d/\delta$ gives $T=\pi/\beta^2$.
Maintaining that $\beta=8$ for the Taylor--Couette system determines a period for the oscillation
$T=\pi/8^2\approx 0.05$.
For the velocity amplitude, 
$U_1\,(\nu/d)=U_0$ implies
$U_1=U_0 d/\nu=(d/\delta)(U_0\delta/\nu)=\beta\, R^\delta$.
% $T=\pi/\beta^2$ and $U_0=R^\delta\beta$.
%Maintaining that $\beta=8$ for the Taylor--Couette system determines a period for the oscillation
%$T=\pi/8^2\approx 0.05$ 
The critical $R^\delta=86$ predicts that the amplitude is critical for $U_1=688$. 
\AW{The most unstable wavenumber $\alpha=0.5$ implies} $m=2\pi R_o/\lambda=2\pi(1/(1-\eta))d/((2\pi/\alpha)\delta)$, which for radius ratio $\eta=0.97$ and $\beta=8$ give $m\approx133$. 
Using these parameters for the Taylor--Couette setup,
figure \ref{fig:KD}({\it a}) shows the real part of the three leading growth rates for perturbations to frozen mean profiles.
The range in $t$ shown corresponds to half a period, as the mean flow is reversed 
in the second half and has the same growth rates.
(Nine oscillations are simulated first to eliminate any transient originating from the 
initial condition $\vec{u}'=\vec{0}$ at $t=0$.)
% and $\cos$ is replaced by 
%$\sin$ so that the initial condition matches the boundary condition.) 
The leading growth rate just touches criticality at the expected time in the cycle, indicating that the 
prediction derived from the channel calculation of \cite{von1974linear}
is very good.

Qualitatively, the structure of the growth rates in figure \ref{fig:KD}({\it a}) is not dissimilar to those of figure \ref{fig:Eigvals(t)}({\em c}) for the 
%To make firmer the link the numerical approximation of the 
start-stop experiment of \S\ref{sect:ApproxOfExpt}, 
\AW{
likewise the structure of the eigenfunction of figure \ref{fig:KD}({\it b}) is similar to that of figure \ref{fig:prof2}({\it b}) for the start-stop experiment at the later time, when it grows most rapidly.}
To make an approximately quantitative link,
\AW{and to a posteriori rationalise the choice of parameters for the start-stop experiment,}
we make a rough approximation that the linear
ramp up and ramp down over a time $T_2$
corresponds to one half
of a sinusoidal oscillation of period $T=2\,T_2$.
Then, $T=\pi/\beta^2$ for our experiment with $T_2=0.05$ implies $\beta=\sqrt{\pi/0.1}\approx 5.6$.  This value is not so far from the  $\beta=8$ of \cite{von1974linear}, suggesting that,
if the instability is of the same nature,
the result
of \S\ref{sect:ApproxOfExpt}
should not have been substantially influenced by the inner boundary. It should then be possible to predict the most unstable wavelength of the 
instability from that of the Stokes problem.
The wavenumber $\alpha=0.5$ based on units $\delta$ for the channel 
corresponds to an azimuthal wavelength of
$\lambda=(2\pi/\alpha)/\beta\approx 2.2$ 
in gap widths $d$.  
In figure \ref{fig:Eigvals} it was found that the peak growth rate 
occurred for $\lambda\approx 2.5$ in the start-stop
experiment, 
a difference of around 
10-20\%.
%14\%. \MB{to stay somewhat qualitative here I would say 10-20\%}

%We conclude this section by suggesting therefore that the start-stop and Stokes' Oscillating boundary instabilities are related.

%COMPARE mean profile at $t=0.03-0.04$ from ramp up-down with oscillation
%at $t=\pi/2$. (?)

\subsection{Dependence on radius ratio}

In this section
we return to the start-stop problem and
consider the case where $R_o$ is fixed and $R_i$ decreases, so that the gap $d$ increases and $\eta=R_i/R_o$ decreases.

If $\Omega_o(t)$ is also unchanged,
then the penetration depth $\delta$ does not change, and $\beta=d/\delta$ increases.  It has been confirmed in calculations that the flow near the outer cylinder is independent of the inner cylinder, and there is no change to the instability (therefore not shown).  

If $d$ is increased and $\Omega_o(t)$ scaled such that $\delta$ increases proportionately, i.e.\ $\beta=d/\delta$ is held fixed, then wavelength of the instability will continue to scale with $d$ and the wavenumber $m$ decreases.  In this case we expect more influence from the curvature.  Conveniently, the dimensionless velocity at the boundary, $V(r_o,t)$, and the time-dependent Reynolds number $Re_o(t)$ are unchanged, as our non-dimensionalisation is already based on the viscous time scale and on $d$.  Only the wavenumber $m$ changes with $\eta$:  for a change $d\to\tilde{d}$ and 
$\eta\to\tilde{\eta}$, we have $\tilde{m}=m\, d/\tilde{d}=m\,(1-\eta)/(1-\tilde{\eta})$.

\begin{figure}
  \centerline{
  \begin{picture}(0,0)
  \put(5,125){({\it a})}
  \put(190,125){({\it b})}
  \end{picture} 
     \includegraphics[width=65mm]{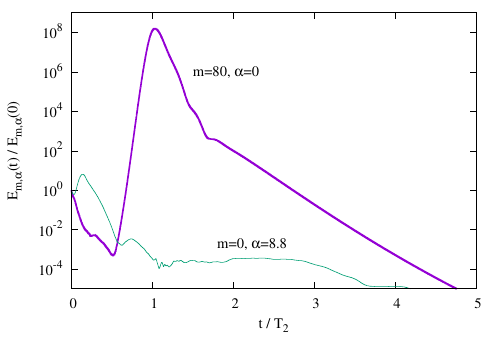}
  \includegraphics[width=65mm]{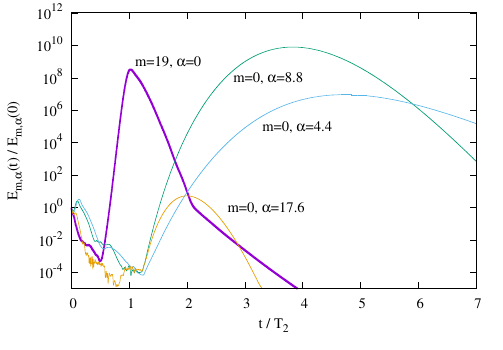} 
  }
  \centerline{
  \begin{picture}(0,0)
  \put(5,140){({\it c})}
  \put(200,140){({\it d})}
  \end{picture} 
  \includegraphics[width=70mm]{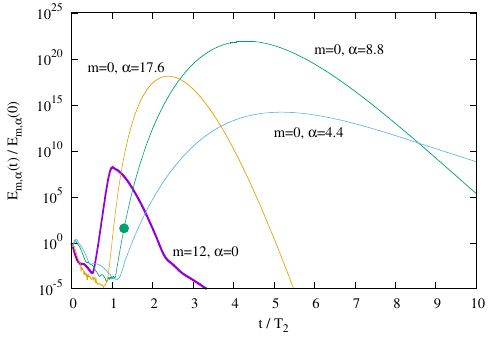}
  ~~~~~
  \includegraphics[width=40mm]{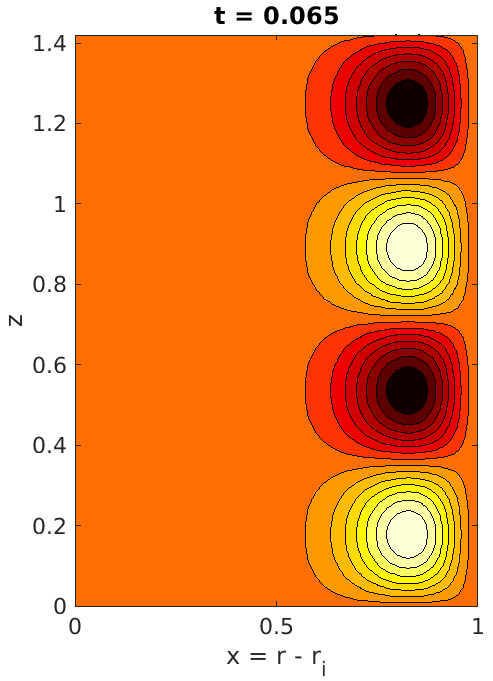}
  }
  \caption{
  \label{fig:Ek1m1}
  Energy of 
  \AW{longitudinal modes ($m\ne0,\,\alpha=0$) and G\"ortler modes ($m=0,\,\alpha\ne0$)} during a start-stop
  with parameters $U_1=8000$, $T_1=0.01$, $T_2=0.05$ at radius ratios
%  modes of azimuthal and axial wavenumbers $m$ and $\alpha$ for 
  ({\it a}) $\eta=0.97$, 
  ({\it b}) $\eta=0.874$,
  ({\it c}) $\eta=0.8$. 
  %with
%  start-stop parameters $U_1=8000$, $T_1=0.01$, $T_2=0.05$.   
  For all $\eta$, the $m\ne0$ for longitudinal modes correspond to azimuthal wavelength $\lambda\approx 2.6$. 
  \\
  \AW{
  ({\it d}) 
   G\"ortler mode at time indicated by the point in ({\it c}); contours of the Stokes streamfunction $\psi$, such that $u_r=-(1/r)\partial_z\psi$. 
   } }
\end{figure}

Figure \ref{fig:Ek1m1} compares the growth of different Fourier modes for $\eta=0.97,~0.874,~0.8$.  
%The initial condition for the radial component of the mode is $u_r(x)=x^4(1-x)$, where $x=r-r_i$, and the $u_\theta$ or $u_z$ component is determined via the continuity condition.
The longitudinal vortices are axially independent, $\alpha=0$, and the non-zero $m$ corresponds to $\lambda\approx 2.6$ for all radius ratios $\eta$. The growth of this longitudinal mode is similar for all cases
\AW{(thick purple line; note the scale on the $y$-axis).}
\AW{Supposing that the penetration depth $\delta$ approximately gives the distance from the outer wall to the centre of a vortex, then it is estimated that the wavelength over a pair of azimuthally independent ($m=0$) G\"ortler rolls
might be $4\,\delta/d=4/\beta$, in units of gap width.}
For $\beta= 5.6$, the 
axial wavenumber is then $\alpha=2.8\,\pi\approx 8.8$.  \AW{In figure \ref{fig:Ek1m1} it is seen that} such G\"ortler modes do not grow for the narrow gap at $\eta=0.97$, but for $\eta=0.8$ they are expected to swamp any trace of the longitudinal vortices in a flash.  
Interestingly, Coles' choice of 
$\eta=0.874$ appears to have been a 
serendipitous %good 
one, 
%\MB{instead of good: prescient?  serendipitous?}, 
close to the borderline where both modes might be observed.

%\subsection{Start-stop versus sudden-stop}
%Most studies of the sudden stop consider the case where the flow is initially in solid body rotation.  \cite{singh2021turbulence} consider a closer case to that here, where the inner cylinder is stationary, and the outer cylinder is stopped suddenly.
%$Re_i=0$, $Re_o=1700$
%Stop {\em very} sudden ~0.0017 visc units.
%
%TRY $T_1=0.01*c$,  $T_2=T_1+0.05/c$.
%NB, $0.05/0.0017\approx 30$.
%MAINTAIN $T_2-T_1=0.04$ but gradually increase $T_1$.

\section{Conclusions}

The figure 22(o) of
\cite{coles1965transition}, captioned
``{\it (o) instability following start-stop motion of the outer cylinder},'' with its longitudinal vortices, has been a bit of a longstanding mystery.  When Coles described his procedure for introducing different azimuthal wavenumbers $m$ for wavy Taylor-vortex rolls, he mentioned a sudden-stop that induced a large number of waves, ``{\it probably as a result of Tollmien instability},'' but referred to the image captioned for a start-stop experiment.

In this work, we have simulated the appearance of longitudinal rolls in the Taylor--Couette system, guided by its unexpected appearance in a more recent experimental campaign \citep{Burin12}.
We have shown that instability is linked to an inflection point in the mean profile (rather than the Tollmien instability) ---  despite the transient nature of the start-stop flow, it is possible to link the instability to that of Stokes oscillating boundary problem, 
\AW{which has a Rayleigh instability for sufficiently large Reynolds numbers 
\citep{cowley1987high}.}

There are a number of reasons why the instability may have remained elusive in  Taylor--Couette flow.  One, of course, is the rich array of other instabilities exhibited by the system, which create other interesting and more manifest diversions.
In the recent experiment, longitudinal vortices were only observed for the case $\eta=0.97$.  
Guided by this experiment, our parameters suggested a ratio for the viscous scale $\delta$ and the gap $d$ of $\beta=d/\delta\approx 5.6$.  From the radial extent of the rolls, seen in figure \ref{fig:prof2}({\it b}), it is expected if $\beta$ is significantly decreased, then without sufficient physical space for the rolls, 
%\MB{see comment in email}, 
the instability will be suppressed.

In our calculations where $\beta$ is fixed while decreasing the radius ratio, i.e.\ the rolls scale with the gap width, the radius ratio $\eta=0.874$ of \cite{coles1965transition} appears to be around the lower limit at which the instability might be observable. At lower $\eta$, 
the magnitude of curvature terms raises relative to viscous terms, and the instability
is almost immediately swamped by 
centrifugal instability and the formation of G\"ortler rolls.

It is less obvious why the instability has not been observed for larger $\beta$, where the gap $d$ widens but $\delta$ is kept the same.
This occurs when $R_o$ and $\Omega_o(t)$ are unchanged and only $R_i$ is reduced.  
Although the instability should be observable in principle, there are practical challenges, and few, if any, of the experiments we have referenced have been looking in the right regime, combining a start-stop or sinusoidal oscillation with a large Reynolds number.
%consideration is the Reynolds number.  
For the traditional case of a stationary outer cylinder and steadily rotating inner cylinder, for $\eta=0.97,\,0.874,\,0.8$ the critical Reynolds numbers are respectively $\Rey_i\approx 239,\,118,\,95$.  Here,
for a change $d\to\tilde{d}$, matching the peak boundary velocity gives
$\tilde{U}_1\,(\nu/\tilde{d})=U_1\,(\nu/d)$,
which implies $\tilde{U}_1=U_1\,\tilde{d}/d=U_1(1-\tilde{\eta})/(1-\eta)$.  For the  $U_1=8000$, inferred from the experiment 
with $\eta=0.97$
and used in most calculations, 
this gives
$\tilde{U}_1\approx 3.4\times 10^4$
at $\eta=0.874$ and 
$\tilde{U}_1\approx 5.3\times 10^4$
at $\eta=0.8$.
These are the peak values for $Re_o(t)$ for the experiment, and while they are achievable, it is  perhaps not usual to be examining linear stability at such large Reynolds numbers.  

Another practical difficulty for the case of large $\beta$ is that, even if the rolls were present, they only occupy a small fraction of the depth that scales like $1/\beta$.  This would lead to only a small contrast for traditional visualisation techniques using crystalline platelets.  

A final question is the source of perturbations necessary to excite the instability.  Endplates for the cylinders naturally break axial independence of the flow.  They may lead to Ekmann vortices, and can aid the formation of azimuthally aligned Taylor-vortex rolls.  
\AW{In both Cole's and our experiment pictured in figure \ref{fig:expts},
turbulence is present at the ends, but often split end plates and other systems are employed to minimise end-effects.}
%There is no obvious counterpart to break the axisymmetry of the flow.  
If the instability can only be excited from very low level disturbances, in combination with the start-stop being only a transient flow, it is quite possible that in many situations it would not reach sufficient amplitude to be observed.

For the future, with laser-based techniques such as particle image velocimetry (PIV) now more prevalent, such diagnostics may aid further experiments, and possibly lead to more frequent unexpected observations, of this instability over a wider range of radius ratios.  Artificial perturbations, such as axially aligned grooves may also enhance the instability, possibly pushing it into regimes where Taylor or G\"ortler rolls would otherwise take over.

\MB{It would also be interesting to revisit whether a signature of the instability may be observed in the presence of turbulence --\cite{verschoof2018periodically} examined turbulent flow with rapid oscillations of the {\em inner} cylinder,
parameterised by the Womersley number 
$\mathrm{\it Wo}=d\,\sqrt{\omega/\nu}$, 
that correspond to values of 
$\beta=\mathrm{\it Wo}/\sqrt{2}$ in the range $10.5$ to $80.8$, commensurate with the instability discussed above.
They observed that
for shorter modulation periods,
the flow responds with a phase delay. Numerical work 
at large Reynolds numbers %Re = 30,000 
has been performed for 
planar Couette flow by \cite{akhtar2022effect} with and without rotation, and \cite{akhtar2024super} for larger amplitudes and more varied waveforms of the oscillation.  They found that modified versions of Stokes' laminar solution persist subject to modified turbulent viscosity, except in the case of cyclonic rotation.
It would be interesting to determine whether turbulent mean profiles become unstable to a form of longitudinal instability at very large Reynolds numbers.
}

%\backsection[Acknowledgements]{Acknowledgements may be included at the end of the paper, before the References section or any appendices. Several anonymous individuals are thanked for contributions to these instructions.}

\backsection[Funding]{This research received no specific grant from any funding agency, commercial or not-for-profit sectors.}

\backsection[Declaration of interests]{The authors report no conflict of interest.}
\noindent

%%%%%%%%%%%%%%%%%%%%%%%%%%%%%%%%%%%%%%%%%%%%%%%%%%%%%%%%%%%%%%%%%%%%%%%%%%%%%%%%%%%%%
%\appendix

%\AW{
%\section{Calculation of multiple eigenvalues using a timestepper.}\label{appA}
%}

%\AW{[(Combined time-integration and Arnoldi method.)  Appendix necessary?]}

%%%%%%%%%%%%%%%%%%%%%%%%%%%%%%%%%%%%%%%%%%%%%%%%%%%%%%%%%%%%%%%%%%%%%%%%%%%%%%%%%%%%%
\bibliographystyle{jfm}
\bibliography{jfm}
\end{document}